\documentclass{elsart}

 \usepackage{graphicx}
 \usepackage{epsfig}

\usepackage{amssymb}

\begin{document}

\begin{frontmatter}



\title{ Unitarity Corrections and Structure Functions\thanksref{FFK}}
\thanks[FFK]{ Based on the talk presented by M.B. Gay Ducati at the  Pan
American   Advanced Studies Institute (PASI2002),  January 7-18, 2002 .}  

\author{M.B. Gay Ducati, M.V.T. Machado }

\address{ Instituto de F\'{\i}sica, Universidade Federal do Rio Grande do
Sul \\  Caixa Postal 15051, 91501-970 Porto Alegre, RS, BRAZIL.
 }

\begin{abstract}
We have studied the color dipole picture for the description of the deep
inelastic process, mainly the structure functions which are driven directly by
the gluon distribution. Estimates for those functions are obtained
using the Glauber-Mueller  dipole cross section  in QCD, encoding the
corrections due to the unitarity effects which are associated with the
saturation phenomenon. Frame invariance is verified in the calculations of the
observables when analysing the experimental data.  \end{abstract}

\end{frontmatter}

\section{Introduction}

In the  kinematical region of small proton momentum fraction $x$, the
gluon is the main parton driving the  behavior of the deep inelastic
quantities. The standard QCD evolution \cite{DGLAP} furnishes a powerlike
growth for the gluon distribution. This result leads,
at first glance, to the unitarity violation at asymptotic energies, requiring
a sort of  control. In the partonic
language, at the infinite momentum frame,  the small $x$ region
corresponds to the high parton density. The latter is connected with
the black disk limit of the proton target and with the parton recombination
phenomenon. These issues can be addressed through a non-linear dynamics beyond
the usual DGLAP formalism. The complete
knowledge about the non-linear dynamical regime plays an important role in the
theoretical description of the reactions in the forthcoming experiments RHIC
and  LHC, where these effects will be enhanced by the high energies or
by the nuclear probes.

The description of DIS in the color dipole picture is somewhat intuitive,
providing with  a simple representation in contrast to the  involved one from
the Breit (infinite momentum) frame. Considering  small values of the Bjorken variable
$x$, the virtual photon fluctuates into a $q\bar{q}$ pair (dipole) with fixed
transverse separation $r$ at large distances upstream of the target, and
interacts in a short time with the proton. More complicated configurations
should be considered for larger transverse size systems, for instance the
photon Fock state $q\bar{q} \,+\,\rm{gluon}$. An immediate consequence from
the lifetime of the pair ($l_c=1/2 m_p x$) to be bigger than the interaction
one is the factorization between the photon wavefunction and the cross section
dipole-proton in the $\gamma^*\,p$ total cross section. The wavefunctions are
perturbatively calculable, namely  through QED for the $q\bar{q}$
configuration \cite{Nikolaev} and  from  QCD for  $q\bar{q}G$ 
\cite{Wusthoff}. The effective dipole cross section should be
modeled and it includes  perturbative and non-perturbative content.
However, since the interaction strength relies only on the configuration of the
interacting system the dipole cross section turns out to be universal and may
be employed in a wide variety of small $x$ processes.

We have taken into account a sound formalism  providing 
the unitarity corrections to the DIS at small $x$,
namely the Glauber-Mueller approach in QCD. It was introduced by A.
Mueller \cite{Mueller90}, who developed the Glauber formalism to study
saturation effects in the quark and gluon distributions in the nucleus
considering  the  heavy onium scattering. Later developments  obtained 
an evolution equation taking into account the unitarity corrections
(perturbative shadowing), generating  a non-linear dynamics which is connected
with higher twist contributions. Its main characteristic is to provide a
theoretical framework for the saturation  effects, relying on the 
multiscattering of the pQCD Pomeron. In this contribution  we report our studies considering  the parton saturation
formalism to  describe  the observables driven by the gluonic
content of the proton in the color dipole picture \cite{GDMMhep}. The inclusive
structure function $F_2$ is calculated, disregarding the  approximations
commonly considered in previous calculations
\cite{Ayala}. The structure functions $F_L$ and
$F_2^{c\bar{c}}$ are also presented  using the Glauber-Mueller
approach and the rest frame in comparison with the experimental data.

\section{ The DIS in the Rest Frame and the Glauber-Mueller Approach}
 \label{Sec2} 
The rest frame physical
picture  is advantageous since the lifetime of the photon fluctuation and
the interaction process are well defined \cite{BrodDelduca}. The more simple
case  is the quark-antiquark state (color dipole), which is the
leading configuration for  small transverse size systems.  The well known
coherence length is expressed as $l_c=1/(2xm_p)$, where $x$ is the Bjorken
variable and $m_p$ the proton mass. An important consequence of this
formulation  is that the photoabsortion cross section can be derived from
the expectation value of the interaction cross section  for the multiparticle
Fock states of the virtual photon weighted by the light-cone wave functions of
these states \cite{Nikolaev}.  That cross section can be cast in the quantum
mechanical factorized  form, \begin{eqnarray}
\sigma^{\gamma^*p}_{T,L}(x,\,Q^2)=\int \,d^2{\bf r}\,\int_0^1\,dz \,\,|\Psi
_{T,L}(z ,\,{\bf r})|^2\,\,\sigma^{\rm{dipole}}(x,z,{\bf r})\,\,, \label{eq8}
\end{eqnarray}

The formulation above is valid even beyond perturbation theory, since it
is determined from the space-time structure of the process. The $\Psi_{T,L}(z
,\,{\bf r})$ are the photon wavefunctions (for transverse $T$, and
longitudinal $L$, polarizations)  describing the pair configuration: $z$
and $1-z$ are the fractions of the photon's light-cone momentum  carried by the
quark and antiquark of the pair, respectively. The transverse separation of
the pair is $\bf{r}$.  The explicit expressions for the wavefunctions are well
known, \begin{eqnarray} \vert \Psi_{T}\,(z,{\bf   r}) \vert^2  &\:=\:&
\frac{6\, \alpha_{\rm{em}}}{4\,\pi^2 } \: \sum_{i}^{n_f}\,e_i^2 \:
\Big\{\,[\,z^2+(1-z)^2\,]\:\varepsilon^2\,K_1^2\,(\varepsilon\, r) 
\,+\, m_q^2\: K_0^2\,(\varepsilon\, r)\,\Big\} \\\label{eq9}
\vert \Psi_{L}\,(z,{\bf   r}) \vert^2 & \:=\:&
\frac{6\, \alpha_{\rm{em}}}{4\,\pi^2 } \:\sum_{i}^{n_f}\,e_i^2 \:
\Big\{\,4\,Q^2\,z^2\,(1-z)^2\,K_0^2\,(\epsilon\, r)\,\Big\}\,. \label{eq10}
\end{eqnarray}

The  auxiliary variable
$\varepsilon^2=z(1-z)Q^2 + m_q^2$, with $m_q$
the light quark mass, and   $K_0$ and $K_1$ are the  Mc Donald
functions of rank zero and one, respectively. The quantity $\sigma^{\rm{dipole}}$ is interpreted as the cross
section of the scattering of the effective dipole with fixed tranverse
separation ${\bf r}$ \cite{Nikolaev}. The most important feature of the
dipole cross section is its universal character, namely it depends only on the
transverse separation ${\bf r}$ of the color dipole. The dependence on the
external probe particle,  i.e., the photon virtuality, relies in the
wavefunctions.   In general, an ans\"{a}tz
for the effective dipole cross section is obtained and the  
process is analized in the impact parameter space.  The main feature of the
current models in the literature is to interpolate the physical regions of
small transverse separations (perturbative QCD picture) and the large
ones (Regge-soft picture). Here we have used  the Glauber-Mueller approach to
determine the dipole cross section, with the advantage of providing the
corrections required by unitarity in an eikonal expansion. For the large $r$
region, we choose to follow a similar procedure from the saturation model
(GBW) \cite{Golec}, namely saturating  the
dipole cross section ($r$-independent constant value).

Now, we shortly present the main results from the Glauber-Mueller approach.
Considering the scattering
amplitude dependent on the usual Mandelstan variables $s$ and $t$, now written
in the impact parameter representation ${\bf b}$,   
\begin{eqnarray}
a(s,{\bf b})\equiv \frac{1}{2\pi}\, \int\,d^2{\bf q}\,\, \rm{e}^{-i\,{\bf
q}.{\bf b}}\, {\rm A}\,(s,t=-q^2)\,.
 \end{eqnarray}
the corresponding total and elastic cross sections (from
Optical theorem) are rewritten in the impact parameter representation (${\bf
b}$) as 
\begin{eqnarray}
\sigma_{tot} = 2\,\int \,d^2{\bf b}\,\,Im \,\,
a(s,{\bf b})\,; \hspace{1cm} \sigma_{el}  =  \int \,d^2{\bf b}\,\, |a(s,{\bf
b})|^2\,, \end{eqnarray}
An important property when treating the scattering in the impact
parameter space is  the simple definition for the unitarity constraint
\cite{Ayala}.  If the real
part of the scattering amplitude vanishes at the high energy limit,
corresponding to small $x$ values, the solution to the that constraint  is
\begin{eqnarray}
a(s,{\bf b})  =  i\,\left[ \,1-\rm{e}^{-\frac{1}{2}\,\Omega\,(s,\,{\bf b})}
\, \right]\,; \hspace{0.6cm}  \sigma_{tot}  =  2\,\int d^2{\bf b}\,
\left[\, 1-\rm{e}^{-\frac{1}{2}\,\Omega\,(s,\,{\bf b})}\,
\, \right]\,, \label{sigeik} \end{eqnarray}
where the opacity $\Omega$ is an arbitrary real function and it should
be determined by a detailed model for the interaction. The opacity function 
has a simple physical interpretation, namely $\rm{e}^{-\Omega}$ corresponds to
the probability that no inelastic scatterings with the target occur. To
provide the connection with  the Glauber formalism, the opacity function
can be written in  the factorized form $\Omega (s,{\bf b})={\Omega}(s)\,S({\bf
b})$, considering  $S({\bf b})$ normalized as $\int d^2{\bf b}\, S({\bf b})=1$ 
(for a detailed discussion, see i.e. \cite{LevNPB}). 

We identify the opacity ${\Omega}(s \approx Q^2/x; {\bf
r})=\sigma^{\rm{nucleon}}(x,{\bf r})$.  The ($q\bar{q}$ pair)
dipole-proton  cross section is  well known \cite{Ayala,LevNPB} and in double logarithmic
approximation (DLA) has the following form  \begin{eqnarray}
\sigma^{q\bar{q}}_{\rm{nucleon}}(x,r)= \frac{\pi^2
\alpha_s(\tilde{Q}^2)}{3}\,r^2\, x\,G(x,\tilde{Q}^2)  \label{dipdla}
\end{eqnarray}
with the $r$-dependent scale $\tilde{Q}^2=r_0^2/r^2$. Considering Eq.
(\ref{dipdla}) one can connect directly the dipole picture with the usual
parton distributions (gluon), since they are solutions of the DGLAP equations.
In our case, we follow the  calculations of Ref. \cite{Ayala,LevNPB} and
consider the effective scale $\tilde{Q}^2=4/r^2$. From the
above expression, we obtain a dipole cross section satisfying the unitarity
constraint and a framework to study the unitarity effects (saturation) in the
gluon DGLAP distribution function. Hence, hereafter we use the Glauber-Mueller
dipole cross section given by \begin{eqnarray}
\sigma_{\rm{dipole}}^{GM}=2\,\int d^2{\bf b}\,
\left(1-\rm{e}^{-\frac{1}{2}\,\sigma^{q\bar{q}}_{\rm{nucleon}}(x,{\bf
r})\,S({\bf b}) }\right)\,. 
\label{dipolgm}
\end{eqnarray}

In order to perform numerical estimates one needs to define the profile function
$S( b)$. This function contains information about the angular  distribution in
the scattering. We have chosen a simple gaussian shape in the impact parameter
space, $S({b})=\frac{A}{\pi\,R^2_A}e^{-b^2/R^2_A}$, where $A$ is the
atomic number and  $R_A$ is the target radius. We will keep this notation
although we are only concerned with the nucleon case. The $R^2_A$ value should
be determined from the data,  ranging between  $5-10$ GeV$^{-2}$ for the
proton case  \cite{Ayala}. Here, we have used the value ( $R^2_A=5$ GeV$^{-2}$)
obtained from a good description of both inclusive structure function and its
derivative \cite{Victorslope}. Such a value corresponds to significative
unitarity corrections to the standard DGLAP input even in the current HERA
kinematics.

In the calculations we have used the GRV94
parametrization \cite{GRV94}: bearing in mind that
$Q^2=4/r^2$, its evolution initial scale  $Q_0^2=0.4$ GeV$^2$ allows 
to scan dipole sizes up to $r_{\rm{cut}}=\frac{2}{Q_0}$ GeV$^{-1}$ (= 0.62 fm).
For recent parametrizations, where $Q_0^2 \sim 1$ GeV$^2$
($r_{\rm{cut}} \approx 0.4$ fm), the uncertainty due to nonperturbative
content in the calculations would increase. An additional advantage is
that GRV94 does not include non-linear effects to the DGLAP evolution  since
it  was obtained from rather large $x$ values, i.e. this
ensures that GRV94 does not include  unitarity corrections in the initial
scale.  To proceed, for the large $r$ region, we choose the following  ansatz:
the gluon distribution is frozen at scale $r_{\rm{cut}}$, namely
$x\,G(x,\,\tilde{Q}^2_{\rm{cut}})$. Then, for the large distance contribution
$r \leq r_{\rm{cut}}$ the gluon distribution reads as 
$x\,G(x, Q^2 \leq Q_0^2)= Q^2/Q^2_0\,x\,G(x,Q^2=Q_0^2)$
leading to the correct behavior $x\,G(x,Q^2) \sim Q^2$ as $Q^2 \rightarrow 0$
and providing the simplest technical procedure.  
\section{Obtaining the Structure Functions}

\subsection{The structure function $F_2$}

First, we perform estimates for the structure function $F_2$ at the rest
frame  considering the Glauber-Mueller dipole cross section \cite{GDMMhep}. The
expression, with the explicit integration limits on photon momentum fraction
$z$ and transverse separation $r$ is,  \begin{eqnarray}
F_2(x,Q^2)\:=\frac{Q^2}{4\,\pi^2\,\alpha_{\rm{em}}}\:\int_0^{\infty}
\!d\,^2{\bf r}\! \int_0^1 \!dz \:  \left( \vert \Psi_{T}\,(z,{\bf      
  r}) \vert ^2 + \vert \Psi_{L}\,(z,{\bf      
  r}) \vert ^2 \right)\: \sigma_{\rm{dipole}}^{GM}(x,{\bf r}^2)\,.\nonumber
\label{f2dip}
\end{eqnarray} 

\begin{figure}[t]
 \centerline{\psfig{file=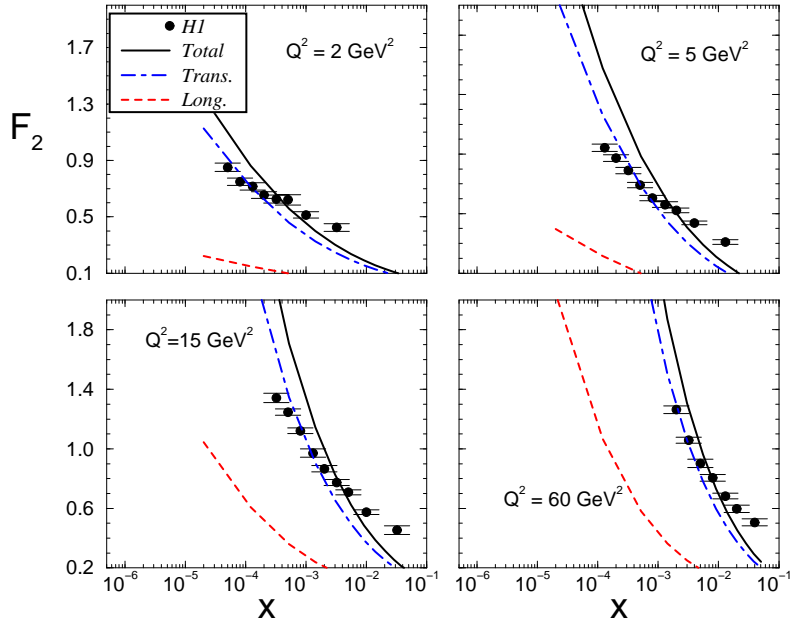, width=100mm}}
\caption{The Glauber-Mueller (GM) result for the $F_2(x,Q^2)$ structure
function. It is shown the transverse contribution (dot-dashed),
the longitudinal one (dashed) and total one (solid line).}
\label{figf2}
\end{figure}

 In the Fig.
(\ref{figf2}) one shows $F_2$ for 
representative virtualities $Q^2$ from the latest H1 Collaboration measurements
\cite{H1f2data}. The longitudinal and transverse contributions are shown
separately. An
effective light quark mass ($u,d,s$ quarks) was taken, with the value
$m_q=300$ MeV, and the  target radius is considered $R^2_A=5$  GeV$^{-2}$. It should be stressed that
this value leads to larger saturation corrections rather than  using radius
ranging over $R^2_A\sim 8-15$ GeV$^{-2}$. The soft contribution comes from
the freezing of the gluon distribution at large transverse separation as
discussed at the previous section. 

\begin{figure}[t]
\centerline{\psfig{file=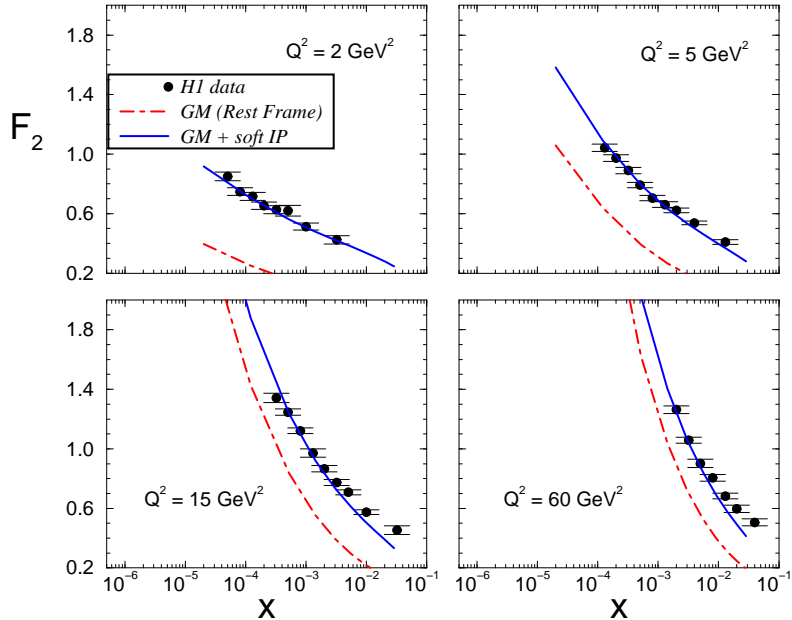, width=100mm}}
\caption{The Glauber-Mueller prediction for the $F_2$ structure function in
the rest frame. For sake of comparison, one uses quark sector ($R^2_A=5$ GeV$^{-2}$, $m_q=0$) and
only transverse wavefunction.  Radius integration $1/Q^2<r^2<1/Q_0^2$ and soft
Pomeron added ( $F_2^{\rm{soft}}={\rm{
C}}_{\rm{soft}}\,x^{-0.08}(1-x)^{10}$).}
\label{f2comp}
\end{figure}

From the plots we verify a good agreement in the normalization, however
the slope seems quite steep. This fact is due to the modelling for the soft
contribution and it suggests that a more suitable nonperturbative input
should be taken.  To clarify the  role played  by the soft nonperturbative contribution to
$F_2$, in the Fig. (\ref{f2comp}) we plot separately the perturbative
contribution and parametrize the soft contribution introducing the
nonperturbative structure function $F_2^{\rm{soft}}={\rm 
C}_{\rm{soft}}\,x^{-0.08}\,(1-x)^{10}$ \cite{Ayalaepjc}, which is added to
the perturbative one. The soft piece normalization is ${\rm
C}_{\rm{soft}}=0.22$.  Accordingly, we have used just
shadowing corrections for the quark sector, taking into account only  the
transverse photon  wavefunction and zero quark mass. The integration on the
transverse separation is taken over  $1/Q^2\leq r^2 \leq 1/Q_0^2$, with
$Q_0^2=0.4$ GeV$^2$ for leading order GRV94 gluon distribution. This leads to
a residual contribution to the soft piece which would come from the transverse
separations $r^2<1/Q^2$. It is again verified that the soft contribution is
important at small virtualities and decreasing as it gets larger. The data
description is quite successful.

Concluding, we have a theoretical estimate, i.e. no fitting procedure,  of the
inclusive structure function $F_2(x,Q^2)$ through the Glauber-Mueller approach
for the dipole cross section, detecting a non negligible importance of a
suitable input for the large dipole size region.

\subsection{The structure function $F_L$}

From QCD theory, the structure function $F_L$ has a non-zero value
due to the gluon radiation, as is encoded in the Altarelli-Martinelli
equation (see \cite{PRD59}), considering the Breit frame.  Experimentally, the
determination of the $F_L$ is quite limited, providing  few data points. Most
recently, the H1 Collaboration has determined the longitudinal structure
function through the reduced double differential cross section, where the data
points were obtained consistently with the previous measurements, however
being more precise and lying into a broader kinematical range \cite{H1f2data}.

\begin{figure}[t]
\centerline{\psfig{file=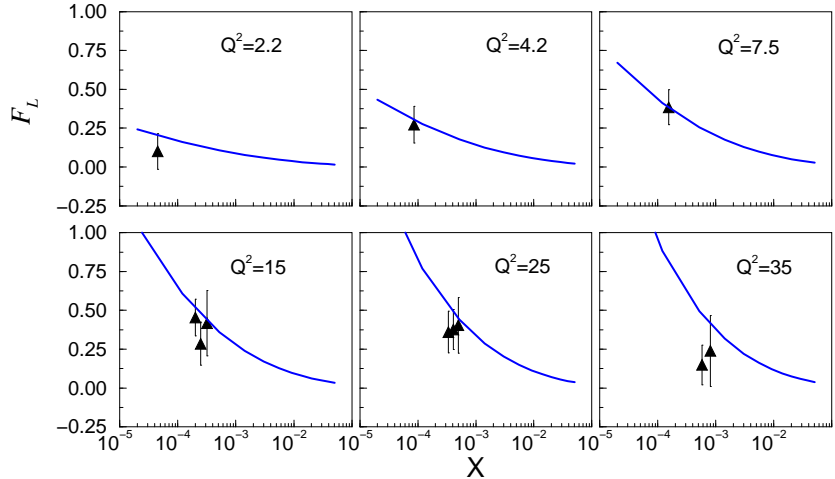, width=100mm}}
\caption{The Glauber-Mueller estimates for the $F_L$ structure function.
One uses light quarks ($m_{q}=300$ MeV), target size $R^2_A=5$ GeV$^{-2}$ and
frozen gluon distibution at large $r$.  Data from H1 Collaboration \cite{H1f2data}.}
\label{flrframe}
\end{figure}

In Fig. (\ref{flrframe}) we present the estimates for the $F_L$ structure
function, in representative virtualities as a function of $x$ \cite{GDMMhep}.   For the
calculations, it was considered light quarks ($u,\,d,\,s$) with effective mass
$m_{q}=300$ MeV and  the target radius $R^2_A=5$ GeV$^{-2}$. The large $r$
region is considered by the freezing of the gluon distribution in this
region. Our expression for the observable is then, 
\begin{eqnarray}
F_L(x,Q^2)\:=\frac{Q^2}{4\,\pi^2\,\alpha_{\rm{em}}}\:\int_0^{\infty}
\!d\,^2{\bf r}\! \int_0^1 \!dz \:   \vert \Psi_{L}\,(z,{\bf         r}) \vert
^2 \: \sigma_{\rm{dipole}}^{GM}(x,{\bf r}^2)\,. \nonumber
\end{eqnarray}

The behavior is quite consistent with the experimental result, either in shape
as in normalization. The quantity is less sensitive to the non-perturbative
content than $F_2$. A better description can be obtained by fine tunning  the
target size or the considered gluon distribution function, however it should be
stressed that the present prediction is parameter-free and determined using
the dipole picture taking into account unitarity (saturation) effects in the
effective dipole cross section \cite{GDMMhep}.  We verify that the rest frame
calculation, taking into account the dipole degrees of freedom and unitarity
effects produces similar conclusions to those ones using the Breit system. For
instance, in a previous work \cite{PRD59}, the unitarity corrections to the
longitudinal structure function were estimated in the laboratory frame
considering the Altarelli-Martinelli equation, with unitarized expressions for
$F_2$ and $xG(x,Q^2)$, obtaining that the  expected corrections reach  to  
$ 70$ \% as $\ln (1/x)=15$, namely on the kinematical corner of the upcoming
THERA project.

The higher twist corrections to the longitudinal structure function
have been pointed out. For instance, Bartels et al. \cite{BGP} have calculated
numerically the twist-four correction obtaining that they are large for  $F_T$
and $F_L$, however having opposite signs. This fact leads to  remaining small
effects to the inclusive structure function by almost complete cancellation
between those contributions. The higher twist content is analyzed considering
the model \cite{Golec} as initial condition. Concerning $F_L$, it was found that the
twist-four correction is large and has negative signal, concluding that a
leading twist analysis of $F_L$  is unreliable for high $Q^2$ and not too small
$x$. The results are in agreement with the simple parametrization for higher
twist (HT) studied by the MRST group in Ref. \cite{MRSTHT}, where
$F_2^{HT}(x,Q^2)=F_2^{LT}(x,Q^2)(1+\frac{D_2^{HT}(x)}{Q^2})$. The second term
would parametrize the higher twist content. In our case, the unitarity
corrections provide an important amount of higher twist content, namely it
takes into account some of the several graphs determining the 
twist expansion.

\subsection{The structure function $F_2^{c\bar{c}}$}

In perturbative QCD, the heavy quark production in electron-proton
interaction occurs basicaly through photon-gluon fusion, in which the emitted
photon interacts with a gluon from the proton generating a quark-antiquark
pair. Therefore, the heavy quark production allows to determine the gluon
distribution and the amount of unitarity  (saturation) effects for the
observable. In particular, charmed mesons have been measured in deep-inelastic
at HERA and the  corresponding structure function $F_2^{c\bar{c}}(x,Q^2)$ is
defined from the differential cross section for the $c\bar{c}$ pair production.

Experimentally, the  measurements of the charm structure function are
obtained by measuring mesons $D^{*\,\pm}$ production \cite{ZEUScharm}.  The function $F_2^{c\bar{c}}(x,Q^2)$  shows an
increase with decreasing $x$ at constant values of $Q^2$, whereas the rise
becomes sharper at higher virtualities. The data are consistent with the NLO
DGLAP calculations. Concerning the ratio $R^{c\bar{c}}=F_2^{c\bar{c}}/F_2$, the
charm contribution to $F_2$ grows steeply as $x$ diminishes. It contributes
less than 10\% at low $Q^2$ and reaches  about 30 \% for $Q^2>120$ GeV$^2$
\cite{ZEUScharm}.

Once more the color dipole picture will provide a quite simple description
for the charm structure function in a factorized way. Now, the Glauber-Mueller
dipole cross section is weighted by the photon wavefunction constituted  by a
$c\bar{c}$ pair with mass $m_c$.  Our expression for the charmed contribution
in deep inelastic is thus  written as 
 \begin{eqnarray}
F_2^{c\bar{c}}(x,Q^2)\:=\frac{Q^2}{4\,\pi^2\,\alpha_{\rm{em}}}\:\int_0^{\infty} \!d\,^2{\bf r}\! \int_0^1 \!dz \:  \left( \vert \Psi^{c\bar{c}}_{T}\,(z,{\bf         r}) \vert ^2 + \vert \Psi^{c\bar{c}}_{L}\,(z,{\bf  r}) \vert ^2 \right)\: \sigma_{\rm{dipole}}^{GM}(x,{\bf r}^2) \nonumber \end{eqnarray}  
where $\vert \Psi^{c\bar{c}}_{T,\,L}\,(z,{\bf      
  r}) \vert ^2$ is the probability to find in the photon the $c\bar{c}$ colo
dipole with the charmed quark carrying fraction $z$ of the photon's light-cone
momentum with $T,\,L$ polarizations. For the correspondent wavefunctions, the
quark mass in Eqs. (2,3) should be substituted by the charm
quark  mass $m_c$. Here, we should take care of the connection between the
Regge parameter $x=(W^2+Q^2)/(Q^2+ 4\,m^2_q)$ and the Bjorken variable
$x_{\rm{Bj}}$. For calculations with the light quarks these variables are
equivalent, however for heavier quarks the correct relation is
 $ x_{\rm{Bj}}=x\,(\, Q^2/Q^2 + 4\,m_c^2\, )$, e.g. see \cite{NikZoller}:.

\begin{figure}[t]
\centerline{\psfig{file=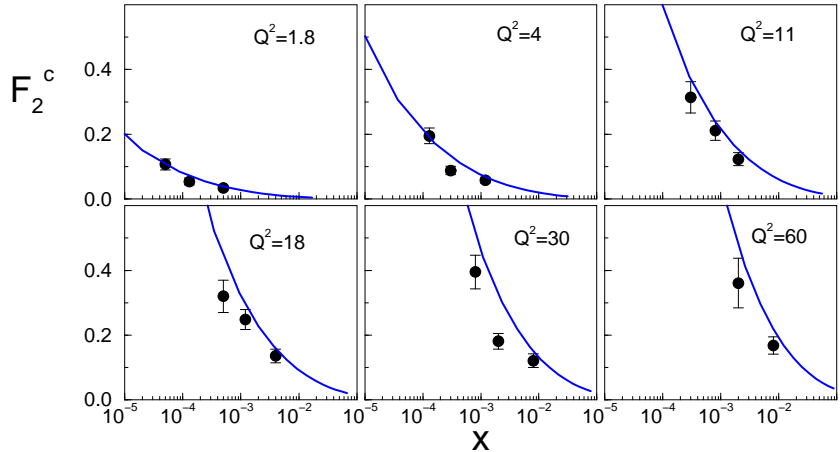, width=100mm}}
\caption{The Glauber-Mueller result for the $F_2^{c\bar{c}}$ structure
function as a function of Bjorken variable $x$ at fixed virtualities (in
GeV$^2$). One uses charm mass $m_{c}=1.5$ GeV, target size $R^2_A=5$ GeV$^{-2}$
and frozen gluon distibution at large $r$.  Data from ZEUS Collaboration
\cite{ZEUScharm} (statistical errors only). }
\label{f2cfig}
\end{figure}

In Fig. (\ref{f2cfig}) we show the estimates for the charm structure function
as a function of $x_{\rm{Bj}}$  at representative
virtualities \cite{GDMMhep}. In our calculations, it was used charm mass $m_c=1.5$ GeV,
target size $R^2_A=5$ GeV$^{-2}$ and frozen gluon distibution at large $r$. We
have verified small soft contribution, decreasing as the virtuality rises.
There is a slight sensitivity to the value for the charm mass, increasing the
overall normalization as $m_c$ diminishes. Such a feature  suggests that  the
charm mass is a hard scale  suppressing  the non-perturbative contribution to
the corresponding cross section. This conclusion is in agreement with the
recent BFKL color dipole calculations of Nikolaev-Zoller \cite{NikZoller} and
those from Donnachie-Dosch \cite{Donnachie}.

Regarding the Breit system description, in Ref. \cite{PRD59} we found
strong corrections to the charm structure function, which are larger than
those of the $F_2$ ones. Considering
the ratio $R_2^c=F_2^{c\,\rm{GM}}(x,Q^2)/F_2^{c\,\rm{DGLAP}}(x,Q^2)$, the
corrections predicted by the Glauber-Mueller approach would reach  62 \%
at values of $\ln (1/x) \approx 15$ (THERA region). Then, an important result
is a large deviation of the standard DGLAP expectations at small $x$ for the
ratio $R^{c\bar{c}}=F_2^{c\bar{c}}/F_2$ due to the saturation phenomena
(unitarization). With our calculation \cite{GDMMhep} one verifies that it is obtained  a good
description of data in both reference systems, suggesting a consistent
estimation of the unitarity effects for that quantity.

\section*{Acknowledgments}
MBGD thanks the organizers for the invitation for this talk and both 
authors enjoyed the  lively  scientific atmosphere of this meeting.

\end{document}